\documentclass{iopart}
\usepackage{amssymb}
\usepackage{graphicx}

\begin{document}

\title{The ``Shoulder'' and the ``Ridge'' in PHENIX}

\author{M P McCumber for the PHENIX Collaboration}

\address{Department of Physics and Astronomy,
SUNY Stony Brook,
Stony Brook, New York, 11794, USA}
\ead{mccumber@skipper.physics.sunysb.edu}
\begin{abstract} The observation of jet quenching in 
ultra-relativistic heavy ion collisions demonstrates 
significant energy loss of fast partons when passing 
through the created medium. Correlations between final-state 
particles at intermediate transverse momentum
(1.0 $\lesssim$ $p_{T}$ $\lesssim$ 4.0 GeV/$c$) allow for 
study of the medium and its response to deposited 
energy. Comparison of these measurements in heavy ion 
collisions with measurements in proton collisions 
show strong modification of the correlation shape and 
particle yields. Two new structures are created, both 
extended in $\Delta\eta$, one centered at $\Delta\phi=0$
(``ridge'') and the other occurring at $\Delta\phi\approx\pi\pm1.1$ (``shoulder''). 
In these proceedings, we describe the measurements of 
these structures that show consistency with a scenario of 
parton-medium interaction and response. We discuss a new 
analysis which selects on the angle of trigger particles 
relative to the reaction plane in Run7. New measurements 
of the centrality and $p_{T}$ dependencies of the structures 
raise the possibility that the same production mechanism 
may give rise to both phenomena.
\end{abstract}
\pacs{25.75.Bh, 13.85.-t}

\section{Introduction}
Jet suppression in heavy ion collisions poses questions about 
the fate of the away-side parton and its lost energy. As 
jet reconstruction is difficult with heavy ion background levels, 
we study jet physics via two particle correlations where the backgrounds are
subtracted statistically\cite{ppg032}\cite{ppg067}\cite{ppg083}.
We refer the reader to the descriptions of this analysis methodology within \cite{ppg083}.

\begin{figure}
\centering
\includegraphics[width=0.95\linewidth]{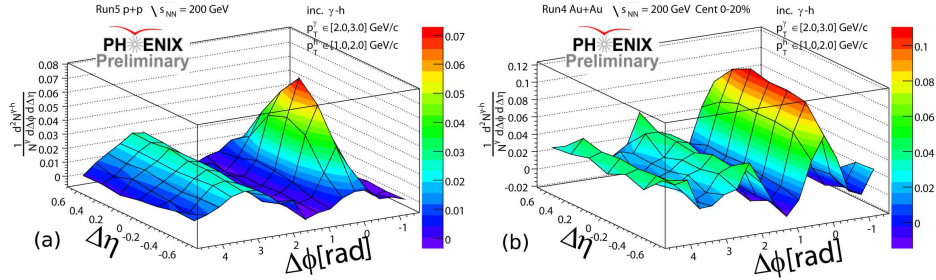}
\centering
\caption{\label{fig:CFs}(Color online) Per trigger conditional yields 
for inclusive photons, $2<p_T^\gamma<3$ GeV/$c$, paired with charged hadron 
partners, $1<p_T^h<2$ GeV/$c$, in p+p and Au+Au collisions, (a) and (b) respectively.}
\end{figure}

In p-p and d-Au collisions we find back to back 
production of jets, but in central heavy ion collisions we 
have the development of two new structures, one on the 
near-side and another on the away-side\cite{ppg083}\cite{ppg039}.
We demonstrate these findings in Fig.\ref{fig:CFs} where we have measured these 
correlations in both the azimuthal angular difference, $\Delta\phi$, and 
the pseudo-rapidity difference, $\Delta\eta$, between two particles. 
In p-p collisions, shown in Fig.\ref{fig:CFs}(a), we see the 
near-side jet, narrow in $\Delta\eta$, and the away-side 
jet, broad in $\Delta\eta$ due to the swing of the away-side 
parton. In contrast, the near-side correlations in heavy ion collisions, shown in Fig.\ref{fig:CFs}(b), 
broaden in $\Delta\eta$. This structure is referred to as the ``ridge''. 
The away-side jet is broadened in azimuth with a yield depletion around $\Delta\phi=\pi$ (hereafter called the ``head'') and a yield enhancement at $\Delta\phi=\pi\pm1.1$ (hereafter called the ``shoulder'').
We conjecture that the head is often dominated by jet suppression and the shoulder by medium
response. Therefore separation of these regions allows study of the interaction between partons and the
medium.

\section{Away-side Modification}
We have in PHENIX two methods for decomposing the 
away-side structure to isolate the characteristics of 
the shoulder mechanism. The first, the bin method, has 
the advantage of being model-independent. This method 
measures the physics of the dominant contribution inside 
a fixed bin in $\Delta\phi$ and suffers in some cases due 
to production from multiple sources. The other, the fit method, decomposes the 
away-side via a multi-Gaussian fit of the form:

\begin{equation}
J(\Delta\phi) = A_{N} e^{\frac{\Delta\phi^{2}}{2\sigma^2_{N}}} + \frac{A_{S}}{2} \left( e^{\frac{\left(\Delta\phi+D\right)^2}{2\sigma^2_{S}}} + e^{\frac{\left(\Delta\phi-D\right)^2}{2\sigma^2_{S}}} \right) + A_{H}e^{\frac{\left(\Delta\phi-\pi\right)^{2}}{2\sigma^2_{H}}}.
\end{equation}
This method has the disadvantage of being model-dependent, but 
should better separate contributions from multiple sources.

A extensive study of away-side correlations as a function of $p_{T}$ selection 
has been carried out \cite{ppg083}. At high $p_{T}$
the away-side peak is similar in shape but suppressed in yield compared to p-p
collisions. Below $p_{T} \approx$ 4 GeV/$c$ the away-side broadens with
the head being suppressed in yield and the shoulder showing yield
enhancement. The latter is found to be more prominent as the associated particle $p_{T}$
decreases \cite{ppg083}.

Measurements of the shoulder $\Delta\phi$ maximum have been made via the fit method 
and show its location is largely independent of both partner and trigger $p_{T}$ selections\cite{ppg083}. 
This disfavors theoretical models which predict a $p_{T}$ dependent location.

We have also measured the away-side particle species dependence.
We find that the correlation shape is similar for both meson and baryon partners. 
The baryon to meson ratio of per trigger associated yields increases 
from peripheral to central collisions. This trend is incompatible with in-vacuum 
fragmentation, but similar to the behavior of the inclusive baryon and meson yields.
Furthermore, we have measured the away-side spectral slopes. Above $N_{part}$ of 100 
where the shoulder mechanism dominates, we find 
that the spectra are softer than in p-p collisions. 
The spectra also show the same $p_{T}$ independence 
that has been measured in other shoulder properties but 
contrasts the expectation from p-p measurements\cite{ppg083}.

\begin{figure}
\centering
\includegraphics[width=1.0\linewidth]{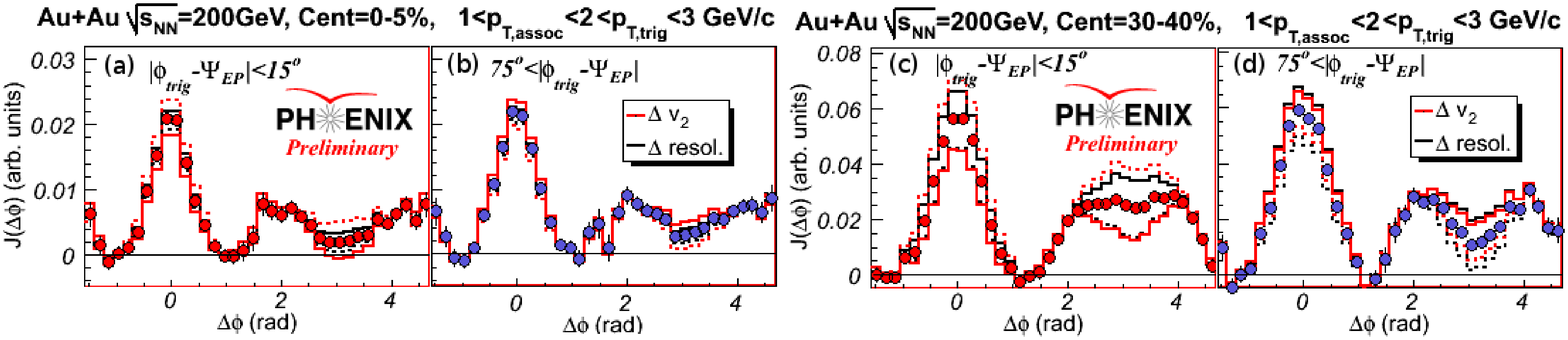}
\centering
\caption{\label{fig:RXPN}(Color online) Jet functions with charged hadron 
triggers, $2<p_{T,trig}<3$ GeV/$c$, paired with charged hadron partners, 
$1<p_{T,assoc}<2$ GeV/$c$. Central collisions, 0-5\%, are shown in (a) \& 
(b), in-plane and out-of-plane respectively.  Mid-central collisions, 
30-40\%, shown in (c) \& (d), again in-plane and out-of-plane respectively.  
Systematic errors are correlated or anti-correlated, as indicated by solid and dashed lines.}
\end{figure}

Path length dependence to the away-side shoulder was probed 
by binning the triggers with respect 
to the reaction plane using the new Reaction Plane detector installed in Run7. 
In Fig. \ref{fig:RXPN} we examine 
a case of little geometry variation in central 0-5\% 
collisions and find no dependence with reaction plane. When we examine 
a case of large geometry variation at 30-40\% centrality, 
we find a slight variation with reaction plane, but this 
may be covered by the current uncertainties. The anti-correlating 
systematic errors are influential in trend determination. 
As understanding of the new Reaction Plane detector improves, this key source of 
systematic error should improve.

\section{Near-side Correlations}
In the near-side, we have measured the per trigger yield with respect 
to p-p collisions via the ratio $I_{aa}$.
We find that the near-side yield enhancement also appears 
in the same $p_{T}$ range, $\lesssim$ 4 GeV/$c$, as the away-side enhancement. The 
highest $p_{T}$ selections again show a similarity to p-p, 
though now without suppression\cite{ppg083}.

We have also measured the near-side baryon to meson ratios and find
that like the away-side the trends move from p-p values 
towards the inclusive measurements. Since the near-side 
jet is not suppressed, the ratios can not achieve an 
inclusive-like particle mixture in the most central 
events\cite{ppg034}. Softening of spectral slopes relative to p-p is 
witnessed for the near-side, again similar to findings 
in the away-side.

We have measured the near-side distributions projected along 
$\Delta\eta$ compared to p-p. This verifies that the near-side 
enhancement measured is broad within our acceptance. Higher $p_{T}$ selections 
match the $\Delta\eta$ profile of p-p. As seen in Fig. \ref{fig:CFs}, outside of 0.5 
$|\Delta\eta|$ we have access to a region relatively 
free of p-p like production for a selection of $p_{T}$ ranges\cite{ppg083}.

\section{Ridge-Shoulder Comparison}
\begin{figure}
\begin{minipage}{0.45\linewidth}
\centering
\includegraphics[width=0.9\linewidth]{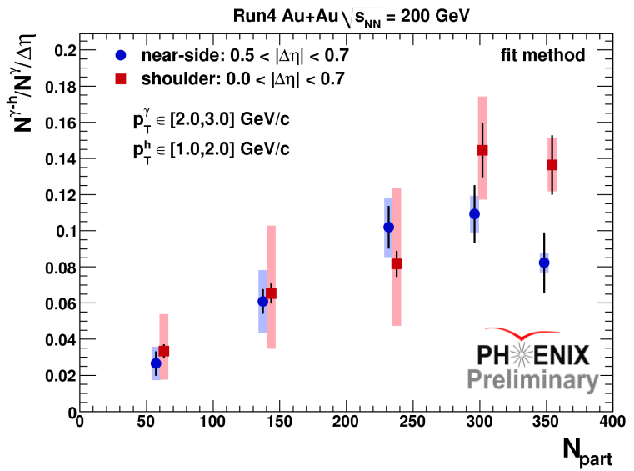}
\centering
\caption{\label{fig:RScent}(Color online) Per trigger conditional yields 
in the ridge-dominated near-side $\Delta\eta$ bin (circles) and in the shoulder
component (squares) by $N_{part}$ in Au+Au collisions.}
\end{minipage}%
\begin{minipage}{0.55\linewidth}
\centering
\includegraphics[width=0.75\linewidth]{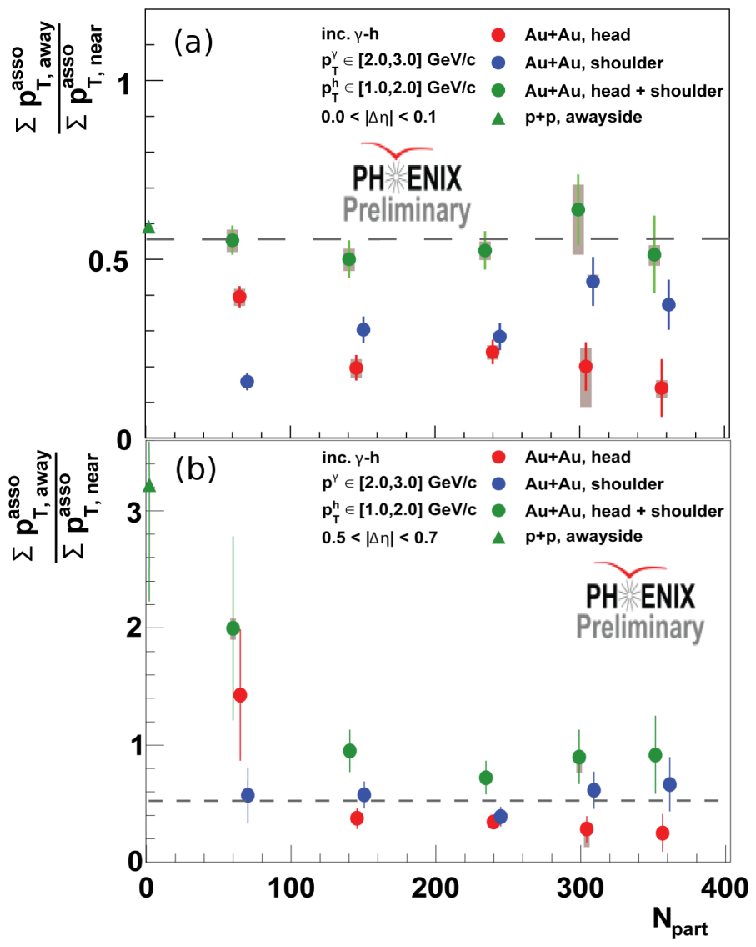}
\centering
\caption{\label{fig:RSbal}(Color online) $p_{T}$ ratios between 
away-side sources and $\Delta\eta$-binned near-side partners}
\end{minipage}
\end{figure}

In this ridge-dominated region, we have 
measured the centrality dependence of the ridge yield 
with respect to the shoulder extracted with 
little contamination from the head (via the fit method).
In Fig. \ref{fig:RScent} we show the ridge and shoulder 
have similar centrality dependencies.

We have also measured the total partner $p_{T}$ in the 
near- and away-side, see Fig. \ref{fig:RSbal}. We find in a 
narrow pseudo-rapidity window, near-side jet and ridge components 
balance with the shoulder and head components. By selecting a 
wide pseudo-rapidity bin, we drop the near-side jet 
contribution and find that the remaining ridge component 
balances with the shoulder component alone. 
Possible independent production of ridge and shoulder leads 
to no expectation that this should be so.
Using this technique we have also measured the 
shoulder and ridge spectra and confirm 
that the spectra are softer than their p-p counterparts. 
The shoulder is closest to the inclusive 
spectral slopes with the ridge being slightly harder\cite{franz}.

The ridge-shoulder similarity found in so many of the 
measurements may be the result of triggering on the 
medium response. If the true shoulder mechanism is 
two-sided and broad in $\Delta\eta$, a trigger arising 
from the medium response would pair with other medium 
response particles at both 0 and $2\pi/3$. We have 
evidence from $I_{aa}$ reported in \cite{ppg083} that the triggers 
below $\lesssim$ 7 GeV/$c$ have significant contributions other
than jet fragmentation. Triggers from jet fragmentation
and medium response would mix ridge and shoulder production (should they be 
produced separate phenomena at all) and could be responsible 
for some of the similarities witnessed in the data.

\section{Conclusions}

PHENIX is performing measurements of both the ridge 
and the shoulder. Both structures are inconsistent 
with in-vacuum jet fragmentation. The ridge and the 
shoulder share much of the same behavior. They 
appear at similar $p_{T}$. They have a similar centrality 
dependence. They are softer than their p-p counterparts. 
They have baryon to meson ratios larger than jet 
fragmentation. And, finally, they balance in $p_{T}$.
These similarities suggest that the ridge and shoulder may
share a common production mechanism.

\end{document}